\documentclass[amsmath,trackchanges]{aastex701}

\begin{document}

\title{Confirmation of the Finch Flatter-Fainter Relation
for the Quadruple Images of Lensed Point Sources}

\author[orcid=0009-0005-3629-0534, gname=Kaitlyn, sname=Roman]{Kaitlyn E. Roman}
\affiliation{The Kavli Institute for Astrophysics and Space Research, Cambridge, MA 02139, USA}
\affiliation{MIT Department of Physics, Cambridge, MA 02139, USA}
\email{ker@mit.edu}

\author[orcid=0000-0002-5665-4172,gname=Paul, sname=Schechter]{Paul L. Schechter}
\affiliation{The Kavli Institute for Astrophysics and Space Research, Cambridge, MA 02139, USA}
\affiliation{MIT Department of Physics, Cambridge, MA 02139, USA}
\email[show]{schech@mit.edu}

\begin{abstract}
  \citet{Finch_2002} derived relations for the summed absolute magnifications of
  quadruply lensed images that vary inversely with flattening in two alternative isothermal gravitational potential models. However, they did not elaborate on the selection effects this ``flatter-fainter'' relation induces in actual lensed systems. We test the relation against the \citet{Luhtaru_2021} sample of 39 quadruply lensed quasars (38 of which we model successfully), using predicted rather than observed magnifications to avoid the complication of microlensing. We found that the summed predicted magnification decreases by a factor of ten over the observed range of flattening.
\end{abstract}


\section{Introduction}

By virtue of exhibiting four images rather than two, quadruply lensed
sources are subject to multiple selection effects.  The larger the
astroidal caustic, the greater the probability that a souce will fall
within its boundaries.  But lens magnifications vary inversely as caustic
area, giving rise to caustic area bias
\citep{baldwin2023malmquistlikebiasinferredareas}.

\citet{Finch_2002} examined the average summed fluxes of the four images
produced by a) the singular isothermal spherical potential elongated
by external shear (henceforth SIS+XS) and b) the singular isothermal
quadrupole potential (henceforth SIQP), given respectively by
\begin{equation} \label{SIS+XS and SIQP}
\psi_{\rm SIS+XS}(r, \theta) = br + \frac{\gamma_{\rm ext}r^2}{2}\cos(2\theta)\quad;\quad \psi_{\rm SIQP}(r, \theta) = br + \epsilon_{\rm QP}br\cos(2\theta) \quad .
\end{equation}
\noindent where $\psi$ is the dimensionless gravitational potential,
$b$ represents the Einstein ring radius (which sets the mass scale),
and $\epsilon_{\rm QP}$ is the ellipticity in the SIQP case.

In what follows, we call the results, summarized in their Table 1 but
not otherwise emphasized, the ``Finch Flatter-Fainter Relations.''
The relations vary inversely as the external shear, $\gamma_{\rm
  ext}$, and inversely as the ellipticity of the quadrupole, in the
SIS+XS and SIQP cases, respectively.  The relations are not valid in
the naked cusp regimes, $\lvert \gamma_{\rm ext} \rvert > \frac{1}{3}$
and $\lvert \epsilon_{\rm QP} \rvert > \frac{1}{5}$.

Our interest in the Finch flatter-fainter relation arose out of our
exploration of the expected fluxes from quadruply lensed supernovae,
the discovery of which with the NSF--DOE Rubin Observatory Legacy
Survey of Space and Time (LSST) is eagerly anticipated
\citep{arendse2024detectingstronglylensedtypeia}. For the sake of
computational speed, the Witt--Wynne model, as described by
\citet{Schechter_2026}, was used. It is based on the singular
isothermal elliptical potential with parallel external shear
(henceforth SIEP+XS$_{\rm \parallel}$):
\begin{equation} \label{SIEP+XSII}
\psi_{\rm SIEP+XS_{\parallel}}(x, y) = b\sqrt{q_{\rm pot}x^2+\frac{1}{q_{\rm pot}}y^2} - \frac{\gamma_{\rm ext}}{2}(x^2-y^2) \quad ,
\end{equation}
which is slightly different from the SIQP in having higher-order even multipoles of the form $\cos(4\theta)$ and higher \citep{baldwin2023malmquistlikebiasinferredareas}. Here we adopt the Finch  et al convention rather than
Baldwin's, elongating the potential along the y-axis.

In what follows, we make use of the semi-ellipticity,
$\eta \equiv (1 - q_{\rm   pot})/(1 + q_{\rm pot})$,
for the elliptical potential. When only
the four image positions are used to constrain a SIEP+XS$_{\rm
  \parallel}$ model, there is a degeneracy between the ellipticity and
the shear, such that the sum of the semi-ellipticity $\eta$ and the
shear $\gamma_{\rm ext}$ is invariant.  The image positions
by themselves cannot separate the flattening of the lensing galaxy by
tidal shear from its neighbors. As shown by \citet{Luhtaru_2021}, this
degeneracy is broken by assuming that the center
of the potential is coincident with the center of the lensing galaxy.

\section{Quadruply Lensed Quasars}

\begin{figure*}[ht!]
\centering
\includegraphics[width=\linewidth]{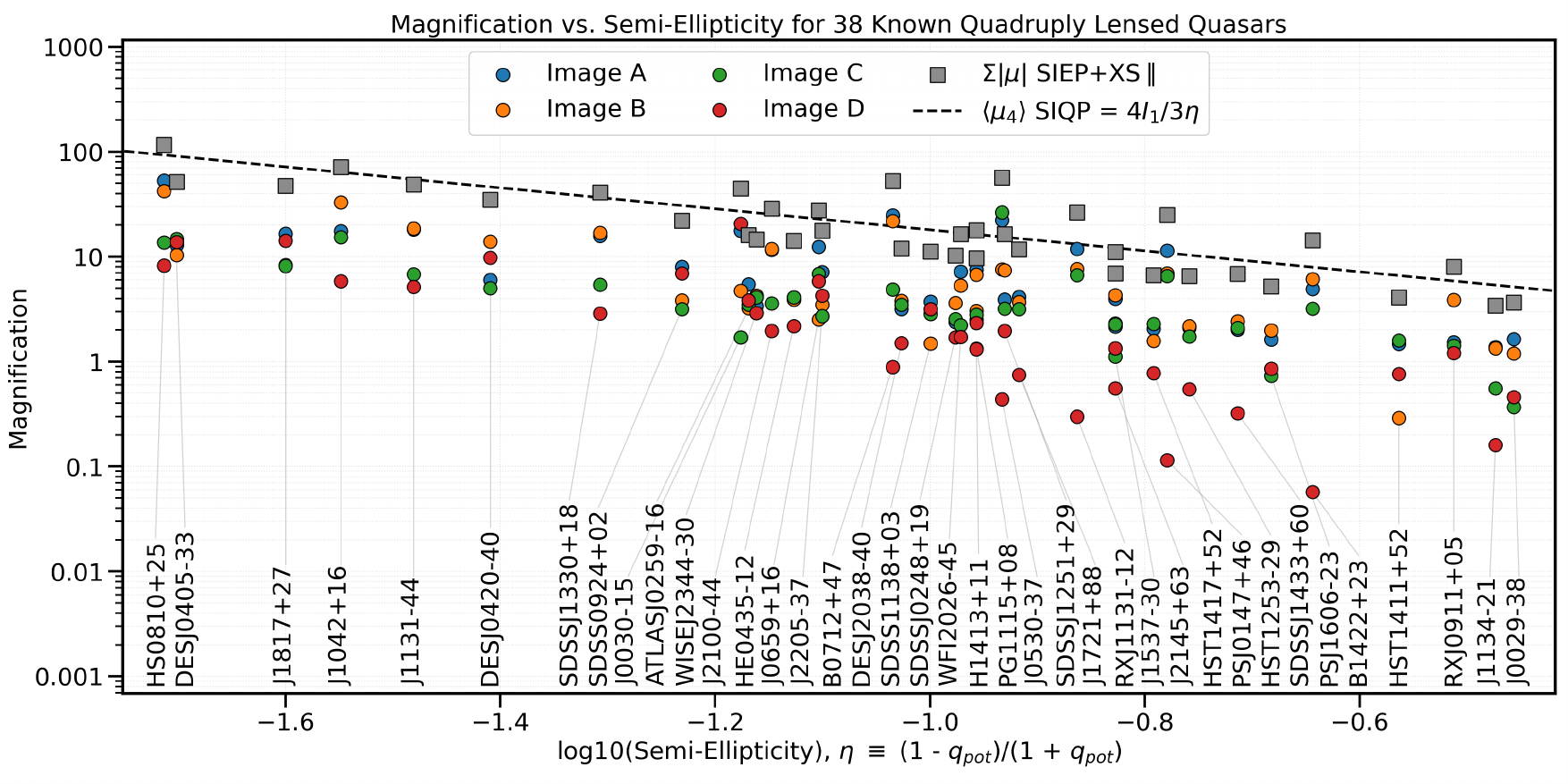}
\caption{A plot of logarithmic semi-ellipticity $\eta$ versus absolute magnification values for 38 known quadruply
lensed quasars. The solid-color circles represent individual image magnifications, while solid-gray squares denote the summed magnification $\Sigma|\mu|$ produced by the Witt--Wynne model. The black dotted line represents the predicted Finch SIQP flatter-fainter relation magnifications $\langle\mu_4\rangle$.}
\label{fig:flatterFAINTER}
\end{figure*}

Rather than simulate configurations, we considered the sample of 39
relatively isolated quadruply lensed quasars assembled by
\citet{Luhtaru_2021}. The predicted absolute magnifications for the four model
image positions are displayed in Figure \ref{fig:flatterFAINTER} as
colored circles. We use the labeling convention of Luhtaru  et al.
A gray square represents the sum of these magnifications.

For the SIQP of Equation~(\ref{SIS+XS and SIQP}), the equipotentials
have axis ratio $(1-\epsilon_{\rm QP})/(1+\epsilon_{\rm QP})$,
which we defined to be the semi-ellipticity, $\eta$.  For the sake of
comparison, we parameterize both models using $\eta$.

Evaluated at the same axis ratio as our SIEP+XS$_\parallel$
models, the \citet{Finch_2002} mean summed magnification for the SIQP
is $\langle\mu_4\rangle = 4I_1/(3\eta)$ for all $\eta$, where $I_1
\approx 1.3511$ is a dimensionless constant obtained by numerical
integration over all source positions.

The system B2045+26 has not been included in Figure
\ref{fig:flatterFAINTER}.  It demands a more complex model than our
SIEP+XS$_{\rm \parallel}$ \citep{McKean_2007}.

The \citet{Finch_2002} predictions, given by $\langle \mu_4 \rangle$
in their Table 1, are shown as a dotted line. Though there is a small
offset of 0.03 dex and rms scatter of about 0.21 dex, the summed
quasar magnifications confirm the Finch et al flatter-fainter
trend. They decrease by a factor of ten as the semi-ellipticity
increases by a factor of ten.

Neither the above-mentioned difference between the SIEP+XS$_{\rm
  \parallel}$ model used here and the SIQP model adopted by Finch et
al nor our approximate fitting method appears to have substantially
altered the flatter-fainter trend.

\section{Consequences for Quadruply Lensed Supernovae}

The LSST will issue alerts for millions of supernovae, only a fraction
of which can be evaluated as lensed candidates. One obvious scheme for
winnowing the alerts is to consider only those in the vicinity of
incipient host galaxies known to be lensed
\citep{arendse2024detectingstronglylensedtypeia}.

Searches for incipient lensed hosts are unavoidably magnitude
limited. An incipient host will be more strongly magnified if the
lens is relatively round. Moreover, rounder lenses produce longer
arcs. Because systems with long arcs are widely used to train
convolutional neural networks to search for lenses, the networks will
favor rounder lenses over flatter ones.

This, in turn, has consequences for lens modeling programs.  Though
a program might place a prior on the flattening of the lens
potential \citep[e.g.,][]{Schmidt_2022}, it may not capture the bias
introduced by the flatter-fainter relation.

While we have used the \citet{Finch_2002} SIQP model, we might alternatively
have used their SIS+XS model, which exhibits a corresponding
flatter-fainter relation and has a predicted magnification that is a
factor of two brighter. The combination of flattening and shear
further complicates lens modeling.

\section{Conclusion}

Using a sample of 39 known quadruply lensed quasars, we have shown that the
\citet{Finch_2002} flatter-fainter relation produces a substantial
effect on the predicted image fluxes. Unless they have
configurations very different from those of our quasars, quadruply
lensed supernovae are likely to exhibit a similar factor of ten effect.

\begin{acknowledgements}

K.E.R. gratefully acknowledges financial support from the MIT Undergraduate Research Opportunities Program via the John Reed Fund.

\end{acknowledgements}

\bibliography{schechter.bib}{}
\bibliographystyle{aasjournalv7}

\end{document}